\documentclass[epsfig,12pt]{article}
\usepackage{epsfig}
\usepackage{graphicx}
\usepackage{array}
\usepackage{color}
\usepackage{bm}
\usepackage{amsmath,amssymb,calc}
\usepackage{braket}
\usepackage{slashed}
\usepackage{bbold}
\usepackage{xcolor}

\newcommand{\beq}{\begin{equation}}   
\newcommand{\eeq}{\end{equation}}
\newcommand{\beqn}{\begin{eqnarray}}   
\newcommand{\eeqn}{\end{eqnarray}}

\newcommand{\pt}{\partial}

\newcommand{\gsim}{\lower.7ex\hbox{$
\;\stackrel{\textstyle>}{\sim}\;$}}
\newcommand{\lsim}{\lower.7ex\hbox{$
\;\stackrel{\textstyle<}{\sim}\;$}}
\usepackage{showlabels}

\begin{document}

\begin{flushright}
FTPI-MINN-22-32, UMN-TH-4135/22
\end{flushright}

\begin{center}
{  \large \bf Infrared Renormalons in Supersymmetric Theories}
 
\end{center}

\begin{center}
{\Large Mikhail Shifman} 

\vspace{3mm}

{\it  William I. Fine Theoretical Physics Institute,
University of Minnesota,
Minneapolis, MN 55455, USA}
\end{center}

\vspace{2mm}

\begin{center}
{\large\bf Abstract}
\end{center}

{\small I argue that in large-$N$ supersymmetric QCD infrared renormalons are absent in the conformal window, there is no need in conspiracy, and  
vacuum expectation values of at least some gluon operators vanish. Basing on this conclusion I conjecture that 
 in  supersymmetric gluodynamics (supersymmetric Yang-Mills theory without matter) at least the leading renormalon ambiguity disappears
 which would be consistent with the fact that the gluon condensate vanishes in this theory,  $\langle G_{\mu\nu}^a G^{\mu\nu\,a}\rangle =0$.
}

\section{Introduction}
\label{intro}

Renormalons are presented by a special class of graphs -- the so called bubble chains -- which was identified in 1977 \cite{renormalon}  (see also the review paper \cite{Beneke}) as the source of a factorial divergence
of perturbation theory at high orders. 

The literature on this phenomenon is huge. The so-called \"Unsal resurgence program \cite{Unsal} associated with factorial divergence of perturbation theory (PT) proved to be very useful
in many quantum-mechanics problems, in solutions of partial differential equations and in some  asymptotically free (AF) field theories in which 
the running of the coupling constant in the infrared (IR) domain can be frozen in some way. However, in theories with genuinely strong coupling IR regime, such as QCD,  reenormalons formally appear.  In fact, the renormalon-associated factorial explosion in high orders is a spurious effect emerging because 
formal expressions are used beyond their limit of applicability. What does that mean and what's to be done? In this introductory section I briefly explain the answers to these questions.

\begin{figure}
\centerline{\includegraphics[width=5.5cm]{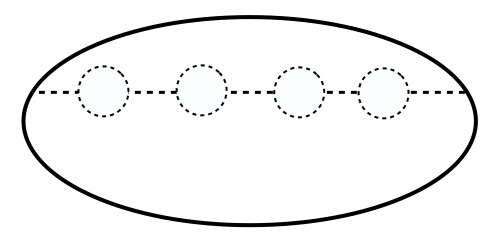}} \caption{\small Graph showing four loops renormalizing a gluon propagator  (represented by the dotted line) attached to the quark loop. A renormalon is the sum 
over all  such diagrams with $n$ loops.
\vspace{2mm} } 
\label{peng}
\end{figure}

Let us consider the diagram shown in Fig. \ref{peng}, representing a typical bubble chain 
in QCD for a leading 
renormalon. The IR contribution of this chain can be written as

\beq
D\propto Q^2 \int
dk^2\frac{k^2\alpha(k^2)}{(k^2+Q^2)^3} \,,
\label{renorm}
\eeq
where $\alpha$ is the running coupling constant and $k^2$ is the momentum running through the long gluon (dashed) line. 
It is obvious that at $k^2\lsim \Lambda^2$ Eq. (\ref{renorm}) makes no sense because $\alpha (k^2)$ does not exist for such values of $k^2$;
quarks and gluons do not exist either. It is straightforward to assess the  corresponding ambiguity, it is  of the order of $O\left(\Lambda^4/Q^4\right)$.

\vspace{2mm}

In the renormalon construction the factorial divergence in (\ref{renorm}) is ``demonstrated" in a completely formal way through the expansion 
of the running
$\alpha(k^2)$,
\beq 
\alpha (k^2) = \frac{\alpha (Q^2)}{1-b_1\frac{\alpha (Q^2)\rule{0mm}{4mm}}{4\pi} 
\ln (Q^2/k^2)} \,,
\label{eight}
\eeq
with the subsequent expansion of the right-hand side  in $\alpha(Q^2)$.
Here $b_1$ is the first coefficient of the $\beta$ function.
Equation (\ref{eight}) describes the running constant $\alpha (k^2)$  expressed in terms of  $\alpha (Q^2)$ with a fixed (large) $Q^2$.
Expanding the denominator  in (\ref{eight}) in powers $\alpha (Q^2)$ we  arrive at the series
\beq
  D (Q^2)\propto \frac{1}{Q^4} \,\alpha \, \sum_{n=0}^\infty \left(\frac{b_1\alpha}{4\pi} \right)^n  \int\,dk^2 \,k^2 \left(\ln \frac{Q^2}{k^2}
  \right)^n\,,\qquad
  \alpha \equiv \alpha (Q^2)
  \label{nine}
  \eeq
  which can be rewritten as
   \beq
  D (Q^2)\propto  \frac{\alpha(Q^2)}{2}\,\sum_{n=0}^\infty  \left(\frac{b_1\alpha(Q^2)}{8\pi} \right)^n  \int\,dy \,
    y^n\, e^{-y}\,,\qquad
 y = 2 \ln \frac{Q^2}{k^2}\,.
  \label{ninep}
\eeq
The $y$ integral in (\ref{ninep}) taken from zero to infinity produces $n!$. 

Observe, however, that a characteristic value of $k^2$
saturating the factorial is exponentially suppressed in $n$,
\beq
k^2 \sim Q^2\,\exp \left(-\frac{n}{2}\right) \,\,\,{\rm or}\,\,\, y\sim n\,.
\label{ten}
\eeq
The factorial explodes as we approach the domain  $k^2\lsim\Lambda^2$,  so that the denominator in (\ref{eight}) hits zero.
This is explained in more detail in Refs. \cite{Shifm2,Shifm3}. Thus, the factorial divergence in high orders in the case at hand is just a signature
of illegitamacy of using (\ref{renorm}) at $k^2\sim\Lambda^2$, an artifact.

QCD and similar theories are self-consistent. Therefore, they must take care of their problems. The correct way to treat  
Fig.\! \ref{peng} is using the
Wilson operator product expansion (OPE). To this end one introduces an auxiliary parameter $\mu$. At strong coupling in confining theories 
it is assumed that $\mu\sim c\Lambda$ where a numerical factor $c$ must be choses say $c\sim 3$ or $\sim 4$, i.e. larger than $\Lambda$, but not parametrically larger.
The virtual momenta $k\geq\mu$ are included in the coefficient functions which become well-defined. The contribution coming from $k\leq \mu$, i.e. from the soft domain,
must be referred to the vacuum expectation values  (VEV) of various operators. In the case of the leading renormalon  (\ref{eight}),  (\ref{nine}) this operator is $G_{\mu\nu}^a G^{\mu\nu\,a}$ with the normal dimension four.

In OPE, the renormalon issue becomes a non-problem; it is replaced by 
the so-called {\em conspiracy}. The conspiracy implies that the coefficients  $C_i(\mu)$ which would contain renormalons 
under the formal procedure of tending $\mu\to 0$ must conspire with the gluon operators $O_i(\mu)$ so that the OPE sum
$\sum_i C_i(\mu)\langle O_i(\mu)\rangle$  ($\langle ...\rangle$ denotes VEVs) is well-defined and $\mu$ independent. Thus, rather than focusing on the ``non-problem" of renormalons we must focus on the conspiracy mechanism. This works perfectly in non-supersymmetric theories (see \cite{Shifm2,Shifm3}
and references therein). 

Below, as a shorthand I will refer to renormalon and the would-be factorial explosion in the coefficients $C_i(\mu) $ (emerging if one tries to send $\mu\to 0$ 
without implementing the proper conspiracy) as the {\em renormalon ambiguity}. This term being quite awkward is widely used in the literature.

In supersymmetric Yang-Mills theories the  mechanism of conspiracy hits an obstacle \cite{DSU}, see also \cite{Shifm3} and \cite{Shifm4}. 
Indeed, the VEV $\langle G_{\mu\nu}^a G^{\mu\nu\,a}\rangle =0$   because $G_{\mu\nu}^a G^{\mu\nu\,a}$ is proportional to the trace of the energy-momentum tensor $\theta^\mu_\mu$ (up to an operator proportional to an equation of motion). 
This leaves the leading renormalon in Fig. \ref{peng} without a conspiracy partner. The issue was addressed in pure ${\mathcal N}=1$ super-Yang-Mills theory (SYM) in \cite{DSU} but not conclusively solved. One of the observations made in \cite{DSU} was as follows.

In SYM theory isolation of the bubble chain graphs is quite non-trivial. The standard practice in QCD reduces to isolating of the {\em matter}
bubble chain, which is charcterized by $b_1^{\rm matter}$ rather than $b_1$ (cf. Eq. (\ref{eight})). Isolating the matter bubble chain is easy and unambiguous, unlike the isolation of the gluon bubble chain which cannot be unambiguously defined in the gauge invariant manner. Then, by default, one just replaces 
 $b_1^{\rm matter}\to b_1$.

In SYM theory this strategy does not work because of the absence of matter. In the subsequent paper \cite{SSS} a study of conspiracy in the two-dimensional supersymmetric $O(N)$ sigma model was carried out in which $\langle \theta^\mu_\mu\rangle$ vanishes too. 
This model  is exactly solvable at large 
$N$.\footnote{We had to consider the {\em next-to-leading order} in the $1/N$ expansion since the leading order is trivial due to factorization \cite{Dav,NSVZPR}.}

Unfortunately, the study \cite{SSS} did not shed light on  the issue of conspiracy in four-dimensional SYM for the following reason. 
Unlike 4D SYM, two-dimensional  $O(N)$ has 
a wider set of dim-2 operators available. And these extra operators (which do not reduce to $ \theta^\mu_\mu$) do indeed conspire to cancel the lowest-dimension renormalon  ambiguity.\footnote{ Of course, in this case we can be certain in these cancellations even before isolating renormalons since the exact solution is well defined and has no ambiguities.}

In a bid to advance our understanding  in four
 dimensions I add matter in SYM theory, thus converting it to  ${\mathcal N}=1$ super-QCD (SQCD). 
  The theory to be discussed below has
SU$(N)$ gauge group and $N_f$ massless matter fields (quark and squarks) in the fundamental representation.
I will focus first on the conformal window  in the Seiberg limit \cite{Sei}.
In this limit we tend $N\to\infty$ keeping the 't Hooft coupling fixed \cite{hooft}. The ratio of $N_f/N$ is a parameter which can changed in a range to be specified below. 

\vspace{2mm}

 I argue that in this theory the renormalon factorials $n!$ and renorma\-lon-associated ambiguities do not appear. No conspiracy 
with the gluon operator  $G_{\mu\nu}^a G^{\mu\nu\,a}$ is needed, $\langle G_{\mu\nu}^a G^{\mu\nu\,a}\rangle =0$ is consistent. Vacuum expectation values
 of other gluon operators in OPE are likely vanish too. 
 
 Based on this conclusions I then return to SYM theory without matter and conjecture that there renormalons are absent too (at least, the leading one) and conspiracy is not required.

\vspace{2mm}

The organization of the paper is as follows. In Sec. \ref{pre} I briefly describe SQCD and remind the main known facts.  Section \ref{arg} is central. Here I present my arguments
 that renormalon ambiguities do not develop in the conformal window. In Sec. \ref{conj} I briefly discuss SYM without matter. This theory is believed to be confining, definitely not conformal. 
 However, even in this case it is natural to  hypothesize that $C_i(Q^2)$ is well-defined at least for the leading operator.
Section \ref{concl} summarizes my conclusions. In Appendix I carry out a direct comparison of the Adler functions in the Seiberg dual pairs.
As was expected, they coincide in the IR limit, at $Q^2\to 0$, but differ at $Q^2\neq 0$.

\section{Preliminaries}
\label{pre}

In this section I will discuss some general aspects of SQCD and the 't Hooft (planar) limit \cite{hooft}. Let us start from the latter.

\vspace{2mm}

{\em Preamble}: \hspace{3mm}  The advantages of the $N\to\infty$ limit are as follows. Instantons and similar quasiclassical contributions are completely suppressed. This eliminates exponential terms 
$\sim e^{-S}\sim \exp(-C/\alpha)$. 
Moreover, the number of planar graphs does not grow {\em factorially}  \cite{Nussinov}.
This leaves us with the infrared (IR)  and ultraviolet (UV) renormalons. The UV renormalons do not introduce  ambiguities.

In SQCD 
Seiberg proved \cite{Sei,Sei2} that two distinct supersymmetric theories -- one with the SU$(N)$ gauge group and the other with  SU$(N_f-N)$ plus an extra color-singlet
``meson" superfield with a super-Yukawa coupling are equivalent in the infrared (IR). This is the so-called ``electric-magnetic" duality. 
Moreover, the gauge coupling $\beta$ functions of the both theories, electric and magnetic, are exactly determined by the NSVZ $\beta$ functions \cite{nsvz} in terms of the anomalous dimensions
of the matter fields. Although the anomalous dimensions $\gamma (\alpha)$ are not BPS protected and hence are not exactly calculable their infrared limit 
$\gamma_*$ is obtained from the requirement of vanishing of the numerator in the NSVZ formula. In the electric theory 

\beq
\gamma_* (\alpha) = - \frac{3N-N_f}{N_f}\,.
\label{14p}
\eeq
The point $3N=N_f$ is the upper edge of the conformal window.  The lower edge of the conformal window can be obtained from the dual ``magnetic" theory (see \cite{Sei}) in which\,\footnote{In the dual theory $\gamma_D$ depends not only on $\alpha$ but also on $f$, see Eq. (\ref{20p}) and the discussion 
that follows it.}
\beq
N\to N_f-N\,,\quad \gamma\to \gamma_D \,\,\, {\rm and}\,\,\, (\gamma_D)_* = - \frac{2N_f-3N}{N_f}
\label{15}
\eeq
implying that at the lower edge $\frac 32 N=N_f$. Thus, the  conformal window stretches in the interval
\beq
\frac 3 2 N \leq N_f\leq 3N\,.
\label{16}
\eeq
The UV fixed point is at $\alpha=0$, the UV and IR fixed points coincide at the edges of the conformal window. As was expected, 
the electric theory is weaker near the right edge of the conformal window, while the magnetic theory is weaker near the left edge.

\vspace{2mm}

At $N_f >3N $ the electric theory is infrared free. At $N+2<N_f<\frac 32 N$ its dual partner is infrared free. Thus, in these two domains there are no infrared renormalons, no conspiracy and no VEVs of gluons operators. Of course, in the UV limit they are in the Landau regime and are not self-consistent 
unless embedded in a larger theory. We will not consider SQCD outside the conformal window. Inside the conformal window both electric and magnetic theories flow in the IR to one and the same conformal theory.

\vspace{2mm}

To follow the strategy of non-supersymmetric QCD in what follows we will have to introduce an external source field. In QCD this role is usually  played by a photon. Therefore, we will add a U(1) gauge superfield to the Seiberg model thus combining SQCD and SQED.
The added U(1) field gauge the ``baryon" symmetry,  will act as a source and will not be iterated in loops. The magnetic theory will support a dual U(1).

\section{Arguments in SQCD}
\label{arg}

Let us start from  $N_f$ is close to $3N$, namely
\beq
N_f= 3N(1-\epsilon)\,,\quad 0< \epsilon\ll 1.
\eeq
Then in the electric theory the anomalous dimension at the IR fixed point $\gamma_*$ is
\beq
\gamma_* \approx {- \epsilon}\,,\quad |\gamma_* | \ll 1.
\label{eleven}
\eeq
At large $N$ the anomalous dimension takes the form
\beq
\gamma (\alpha)=- \frac{N\alpha}{2\pi} +\frac 1 2 \left(  \frac{N\alpha}{2\pi}\right)^2 + C\left[-3 \left(  \frac{N\alpha}{2\pi}\right)^2+\frac{N_f}{N} \left(  \frac{N\alpha}{2\pi}\right)^2 \right]+...
\label{12}
\eeq
where $C$ is a numerical constant $\sim 1$, see \cite{KS}.  Note that at the right edge of the conformal window the
term in the square brackets vanishes. 

\begin{figure}
\centerline{\includegraphics[width=7.5cm]{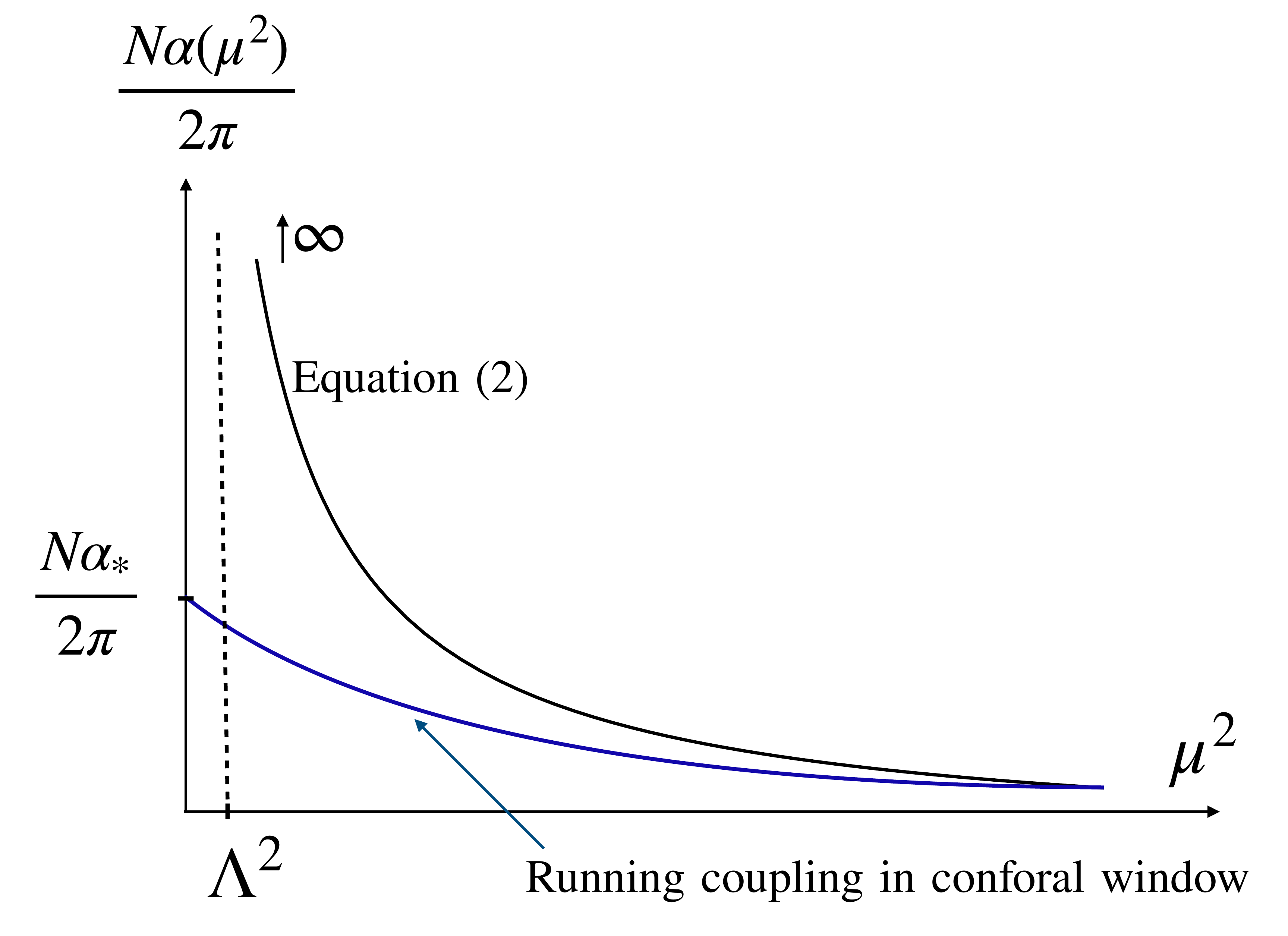}} \caption{\small The running coupling according to Eq. (\ref{eight}) leading to a formal factorial divergence in bubble chain at high orders vs.
the conformal window coupling. Near the edges of the conformal window $\frac{N\alpha_*}{2\pi}\ll 1$  while in the middle $\frac{N\alpha_*}{2\pi} \lsim 1$.
\vspace{-1mm} } 
\label{2c}
\end{figure}

The corresponding IR fixed point for $\alpha$ 
\beq
\frac{N \alpha_*}{2\pi} \approx \epsilon \ll 1\,. 
\label{13}
\eeq
is achieved at $\mu\to0$.
I used only the leading term of Eq. (\ref{12}) in Eq. (\ref{13}).
For the given values of $N_f$ and $N$ the formula (\ref{13}) gives the maximal value of $\alpha$ on the RG flow trajectory on the way from
the UV to IR. Equation (\ref{eight}) is not valid and $N\alpha(k^2)/2\pi$ remains small at small values of $k^2$, as shown in Fig.~\ref{2c}. Moreover,
the approach of $\alpha$ to $ \alpha_*$ in the vicinity of the IR fixed point is power-like, not logarithmic.
Under the circumstances the renormalons do not develop, even formally, the $\alpha$ series must be convergent. This is explained in great detail in the review papers \cite{Shifm2,Shifm3,Shifm4}.   Given that the gauge coupling $\alpha$ is always small in the regime at hand we conclude that gluon operators have vanishing vacuum expectation values. Conspiracy is just not needed.

\begin{figure}[h]
\centerline{\includegraphics[width=8cm]{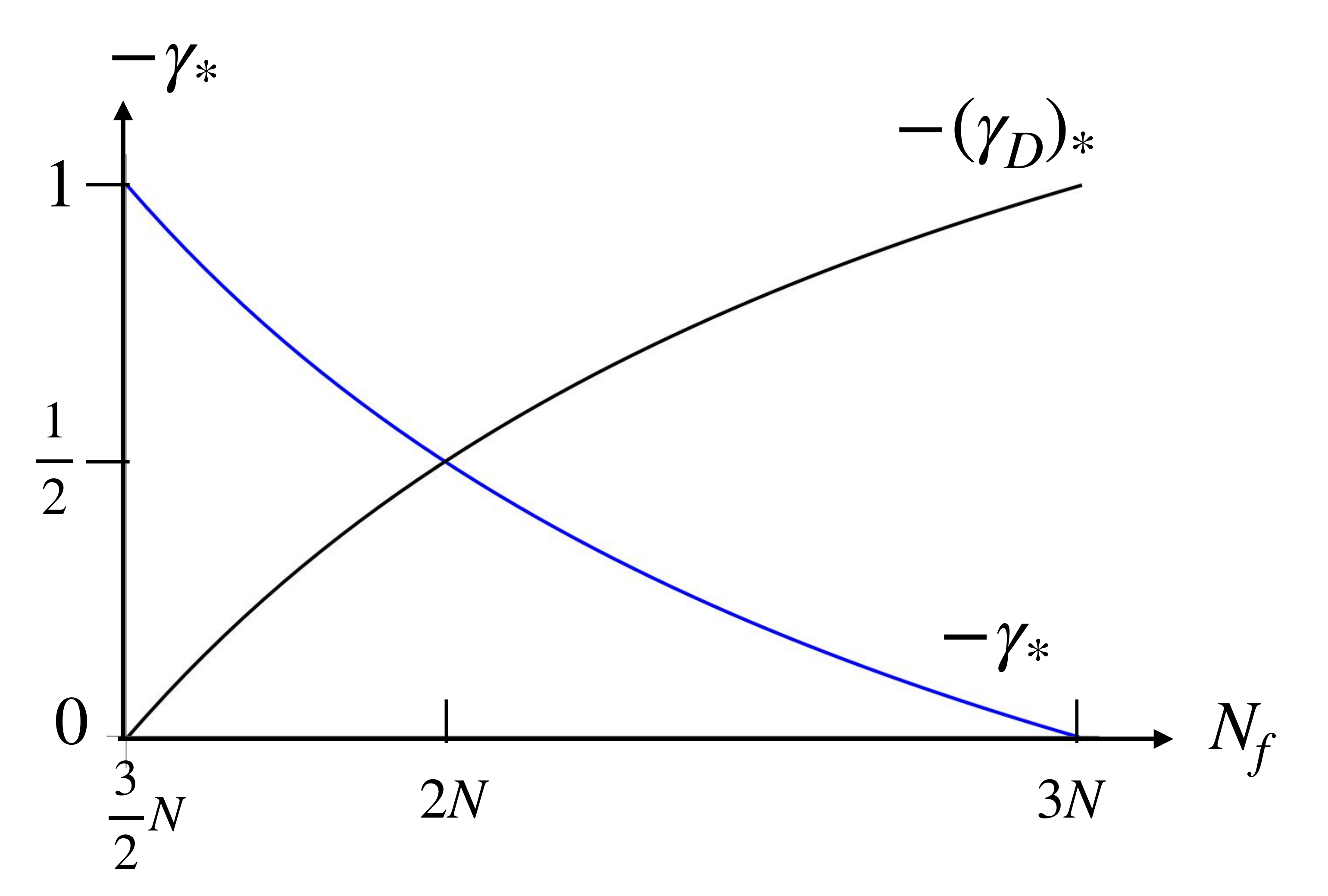}}
\caption{\small 
Anomalous dimensions of matter fields as functions of $N_f$ in the electric and magnetic theories. The magnetic $\gamma$ is marked by the subscript $D$.
} 
\label{FIGU2} 
\end{figure}

Now, let us move toward  smaller values of $N_f$. Then $|\gamma_*|$ increases and at $N_f=2N$ reaches $\frac 12$. At this point one can pass to the dual theory, see Fig. \ref{FIGU2} in which $|(\gamma_D)_*|$ will decrease from $\frac 12$ down to zero as the value of $N_f$ continues to decrease down to $\frac 32 N$ --  but this is not necessary. Even if we continue with the electric theory
the maximal value of $\gamma_*$  achieved at the left edge of the conformal window at $N_f=\frac 32 N$  is $\gamma_*=1$. Then
\beq
\frac{N \alpha_*}{2\pi}\lsim 1\,, 
\label{thirteen}
\eeq
(cf. Eq. (\ref{12})) and $\alpha_*$ does not  explode at small $k^2$ which would be needed for the factorial growth of coefficients to emerge.
This can be proven by analysis of the dual theory. Indeed,  $-\left(\gamma_* +(\gamma_D)_*\right) = 1$ for all $N_f$. Moreover, observable  quantities in the IR in electric and magnetic theories identically coincide, see Appendix.\footnote{If one chose to pass from the electric to magnetic
description at  $N_f=2N,\quad \gamma_*=\frac 12$,
one would see a cusp in Fig. \ref{FIGU2}. This is an artifact. In the ``observable" quantities, such as the Adler functions,
the predictions of the of the electric and magnetic theories identically coincide in the IR for each value of $N_f$ and $N$, provided one replaces
$q_f$ by $\left(q_f\right)_D$. }

An additional argument
in favor of the statement of no renormalons/no conspiracy in all conformal window is that the values of $\gamma_* $ and $(\gamma_D)_* $, see (\ref{14p}), are unambiguous and smooth functions of $N_f$. Everywhere inside the conformal window, we are in one and the same conformal phase,
there are no mass gaps and no irregularities due to opening mass thresholds. 

Summarizing, SQCD in the conformal window exhibits no evidence for renormalon ambiguities which  entails vanishing VEVs of at least some gluon operators.

\vspace{2mm}

Finally I want to make a remark about non-supersym\-metric QCD with massless quarks. This theory also has a conformal window \cite{BZ}. 
 Close to its right edge,\footnote{The exact position of the left edge is still unknown.} i.e. at $N_f\lsim 5.5 N$, the infrared fixed point is at small values of $\alpha$ and, therefore I expect no renormalon ambiguities and no VEVs.
 Unlike SYM, however, when we move to small enough $N_f$ , say, $N_f\sim  N$,  there is a phase transition and, therefore, renormalon ambiguities, conspiracy and 
 OPE with the full set of VEVs re-establish themselves.

\section{Conjectue on supersymmetric Yang-Mills}
\label{conj}

Inspired by the conclusions of the previous section let us ask what happens in pure SYM theory, without matter.
In the infrared this theory is certainly not conformal, rather, it is believed to be confining.
What we know for sure is that the VEV of the leading dimension-4 operator $G_{\mu\nu}^a G^{\mu\nu\,a}$ vanishes. This implies no leading renormalon
ambiguity since ${\mathcal N}=1 $ SYM theory  {\em per se} is unambiguous.

Let us have a look at the NSVZ $\beta$ function in SYM. It can be written as 
\beq
\frac{\pt\left(2\pi/N\alpha\right)}{\pt L}=\frac{3}{1-\tfrac{N\alpha}{2\pi}}\,,\qquad L=\log \mu\,.
\label{msb}
\eeq
Equation (\ref{msb}) is {\em exact}, for more detailed discussion see \cite{KSp} and references therein.
It implies that the  value 
\beq
\frac{N\alpha_*}{2\pi}=1
\label{msbp}
\eeq
is the maximal value of the coupling constant which can be achieved in the $\alpha$ running according to the asymptotic freedom formula.
The value (\ref{msbp}) is approached from below as follows \cite{KSp},
\beq
\alpha_*-\alpha(\mu) \sim {\rm const\,} (\mu-\Lambda)^{1/2},\qquad \mu\gsim\Lambda,
\label{msbpp}
\eeq
see Fig. 1 in \cite{KSp}. Since the regimes (\ref{eight}) and (\ref{msbp})-(\ref{msbpp}) are drastically different it is plausible
that SYM is free from the bubble chain ambiguities.

A related question immediately comes to one's mind: is it possible in principle that {\em confining} super-Yang-Mills theories, as opposed to superconformal,  are compatible with Euclidean OPEs, say, for the two-point functions which have certain VEV's of, say, gluon operators vanishing?
After all, confining SYM theoris must have a mass gap and, at $N\to \infty$, the spectral densities in the two-point functions must looks as a comb built from delta functions at the position of the meson states.

The answer to this question in its most extreme formulation  was found long ago and it was in positive.  In 1978 Migdal   asked himself \cite{Migdal} what is the best possible accuracy to which $\log Q^2$ can be approximated by an infinite sum of infinitely narrow discrete mesons in the spectral function.
The answer is as follows: 

If the mesons are placed at the zeros of a Bessel function, with well-defined residues then no $(\Lambda^2/Q^2)^k$ corrections to  $\log Q^2$
will appear in the Euclidean OPE -- all corrections will be 
 {\em exponentially} suppressed  at large $Q^2$." Much later it was realized \cite{Shifman2005} that just this situation takes place in a holographic QCD model 
suggested in \cite{son}. The authors found that the bare-quark-loop logarithm is represented in their AdS/QCD model as an infinite sum over excited mesons 
on the one hand, and on the other hand the Euclidean OPE is of the form
\beq
\log Q^2+\sum_i a_i \exp \left( - b_i Q^2/\Lambda^2 \right),
\eeq
  -- there are no power corrections at all.
  
  \section{Conclusions}
  \label{concl}
  
  In this paper I revisit a long-standing problem of renormalons in supersymmetric Yang-Mills theories caused by the impossibility of the required  conspiracy in OPE. I argue that in SQCD this problem is absent in the Seiberg conformal window -- there are no reasons for the factorial growth. In SYM without matter the situation may be similar. This latter statement  is a hypothesis.
  
  \section*{Acknowledgements}
  
 I am very grateful to Ken Intriligator and Konstantin Stepanyantz for valuable discussions.
 
 This work is supported in part by DOE grant DE-SC0011842.
 
\section*{Appendix}

In Sec.\! \ref{arg} 
I mentioned that all quantities observable in the IR limit in electric and magnetic theories identically coincide. 
Now I want to demonstrate it in the example of the exact Adler function $D$  relatively recently calculated in  \cite{SS}.
The Adler function is determined by the current-current two-point correlator. In this sense  it can be viewed as
a supersymmetric analog of the study of $e^+e^-$ annihilation to hadrons. 
This result in the IR limit is not new,  it was indirectly obtained in \cite{Anselmi} (see the discussion after Eq. (2.27)), 
where the infrared limit of $D$ was shown to be related by supersymmetry to the triangular  't~Hooft $F\!F\!R$ anomaly which, in turn,  had been matched in dual theories long ago \cite{Sei}. $R$ in $F\!F\!R$ is the anomaly-free $R$ charge.\footnote{ The same relation could be obtained from the papers
\cite{|ntri,Kuta, Erkal} in which  the so-called $a$ maximization was required.  It is worth emphasizing that the result thus obtained is applicable only at the IR fixed points (but generally not along the RG flow away from the fixed points, see my remark at the end of this Appendix).}

\begin{figure}
\centerline{\includegraphics[width=10.5cm]{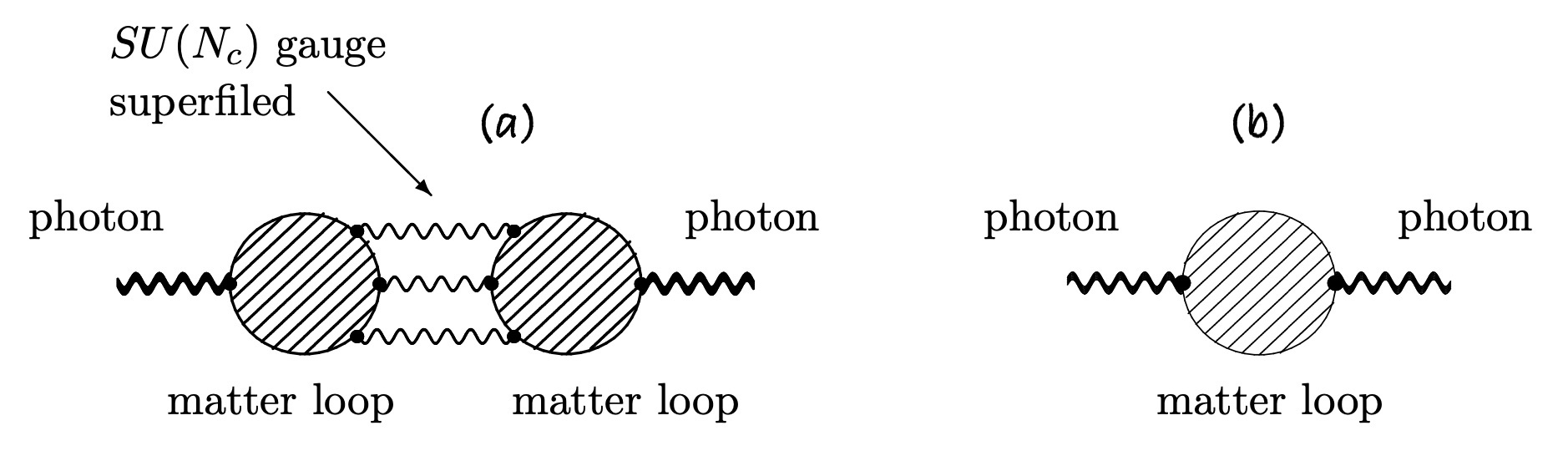}} 
\caption{Graph determining the Adler $D$ functions in Euclidean. Upon analytic continuation to Minkowski  their imaginary part reduces to the total cross-section of $V_\mu \to$ matter. All quasi-diconnected graphs of the type depicted in Fig. \ref{FIGU1}{\em a} vanish due to supersymmetry.} 
\label{FIGU1}
\end{figure}

The demonstration presented below is {\em direct}. 
It  is based on the superfield calculation of graphs presented in Fig. \ref{FIGU1}.

%\marginpar{\tiny R?}

 Since our focus is on strong interactions  we can truncate $e^+e^-$ and consider the two-point function of (virtual) photons, Fig. \ref{FIGU1}. Its imaginary part gives the cross-section for matter production. Thus, we combine SUSY QCD with SUSY QED.  For simplicity, I will assume all electric 
charges of the matter fields to be one, 
\beq
q_f=1\,.
\label{qfe}
\eeq
The exact formula for the Adler function $D(Q^2)$ takes the form \cite{SS},
\beq
D(Q^2) = \frac 32 N \sum_f q_f^2 \Big[1- \gamma (\alpha (Q^2))\Big]
\label{tsix}
\eeq
where the sum runs over all flavors and $\gamma$ is the anomalous dimension of the matter
fields (all matter fields belong to the fundamental representation of color). In our formulation of the problem $\gamma$  is the same for all flavors. 
Equation (\ref{tsix}) is valid for $Q^2\in [0,\,\infty ].$

Supersymmetry implies that all ``semi-connected" graphs of the type presented in Fig. \ref{FIGU1}$a$ vanish \cite{SS}.
Only the graph \ref{FIGU1}{\em b} contributes. Conceptually, our derivation of (\ref{tsix}) was somewhat similar to that of the NSVZ formula \cite{nsvz}. The anomalous dimensions of the matter fields from \cite{nsvz} were used to determine the IR values  $\gamma_*$.  They were also instrumental in the studies of the superconformal $R$-symmetries in four dimensions and their relation to the $a$ theorem \cite{|ntri,Kuta,Erkal}.

 A few words are in order here about the magnetic component of Seiberg's dual pair. 
It contains  an additional  ``meson" color-neutral superfield
${\mathcal M}^i_j$ with a certain superpotential. Hence, in addition to the gauge coupling  a (super-)Yukawa coupling is present too,
\beq
{\mathcal W}_D = f {\mathcal M}^i_j {\mathcal Q} _i   \bar{ {\mathcal Q}}^j\,.
\label{20p}
\eeq
The anomalous dimension 
 $\gamma_D$ depends not only on $g^2_D$ but also on the superpotential coupling constant $f$, and so does the beta function for $f$.
While the superpotential is not renormalized, the coupling $f$ still runs due to the emergence of the $Z$ factors in the matter kinetic terms.
If the number of dual colors (i.e. $N_f-N$) and the number of flavors $N_f$ are large, then
\beq
\beta_f  = - (N_f-N)\frac {\alpha}{2\pi}+ c_f\frac{|f|^2}{4\pi^2}\,,
\eeq
where $c_f$ is a positive number depending on the structure of the matrix superfield ${\mathcal M}^i_j $, for instance $c_f\sim N_f$.
The RG flow of $|f|^2$ toward IR slightly depends on the initial conditions. If we switch off $\alpha$ then 
 $|f|^2$ is IR free. However, the gluon contribution has the opposite effect. Even if at an intermediate scale $\mu$ the constant $\alpha < |f|^2$, under the RG flow toward IR
 $\alpha$ will go up while $|f|^2$ will go down so that  eventually they undergo a crossover; at this point $\alpha$ will become larger and will turn the $|f|^2$ running into asymptotic  freedom type, and then the cycle reverses. In the IR they both hit an IR fixed point which is determined by Eq.~(\ref{15}). On the other hand, starting at point $A$ in Fig. 3 in \cite{Barnes}, one will never see the Landau growth of $|f|^2$ in the IR;  the IR limit is the point C in this figure.  As explained in \cite{Sei2}, the  value of $|f|^2$ at the IR  fixed point can be rescaled by any field rescaling preserving Eq. (5.6) in \cite{Sei2}, where $1/\mu =f$). 
 
\vspace{2mm}

Equation (\ref{14p}) implies that in the infrared limit
\beqn
1- \gamma_* &=& 1 + \frac{3N-N_f}{N_f} =\frac{3N}{N_f}\\[2mm]
\label{5}
&&D_*= \frac{9}{2} N^2\,.
\label{6}
\eeqn
What changes in passing to the magnetic theory? First of all, the U(1) charge becomes 
\beq
q_D =\frac{N}{N_f-N}\,,
\label{qdm}
\eeq
instead of (\ref{qfe}), see \cite{Sei}.
Second, in
 the dual magnetic theory, the number of colors is $N_f-N$. Taking into account these changes   in the IR limit   (see e.g. Eq. (\ref{15}))
\beqn
1- (\gamma_D)_* &=& 1 + \frac{3(N_f-N)-N_f}{N_f} =\frac{3(N_f-N)}{N_f}\\[2mm]
\label{7}
&&D_*= \frac{9}{2} N^2\,,
\label{8}
\eeqn
 where I used Eq. (\ref{qdm}) for the dual U(1) charge. 
We see that the ``observable" total cross-sections in the Seiberg dual pair coincide in the IR for all values of $N_f/N$.

 At the same time, it is quite obvious that at $Q^2\,\, /\!\!\!\!\!\!\to 0$ the electric and magnetic Adler functions differ from each other. Their ratio at the {UV} 
 fixed piont (i.e. $Q^2\to\infty$)
 varies from $\frac 32$ to $3$ depending on the position in the conformal window.

\end{document}